\documentclass[draftclsnofoot, 12pt, onecolumn]{IEEEtran}

\IEEEoverridecommandlockouts    

\usepackage{stmaryrd}
\usepackage{amsfonts}

\usepackage{amssymb}
\usepackage{amsmath}
\usepackage{amsmath,bm}

\usepackage{mathrsfs}
\usepackage{verbatim}
\usepackage{multirow}
\usepackage{array}

	\usepackage[pdftex]{graphicx}
	\DeclareGraphicsExtensions{.pdf,.jpeg,.png}
	\usepackage[pdftex, bookmarks = true, colorlinks=false]{hyperref}
\RequirePackage{lineno}

\usepackage{color}
\usepackage[usenames]{xcolor}

\usepackage{arydshln,leftidx,mathtools}
\usepackage{booktabs}
\usepackage{subfig}

\DeclareMathOperator*{\argmax}{\text{argmax}}

\begin{document}

\setlength{\dashlinegap}{1.0pt}


\title{Enhancing the Accuracy of \\ Device-free Localization Using \\ Spectral Properties of the RSS
}
\author{Ossi Kaltiokallio, H\"{u}seyin~Yi\u{g}itler, and Riku~J\"{a}ntti}


\maketitle

\begin{abstract}
Received signal strength based device-free localization has attracted considerable attention in the research society over the past years to locate and track people who are not carrying any electronic device. Typically, the person is localized using a spatial model that relates the time domain signal strength measurements to the person's position. Alternatively, one could exploit spectral properties of the received signal strength which reflects the rate at which the wireless propagation medium is being altered, an opportunity that has not been exploited in the related literature. In this paper, the power spectral density of the signal strength measurements are related to the person's position and velocity to augment the particle filter based tracking algorithm with an additional measurement. The system performance is evaluated using simulations and validated using experimental data. Compared to a system relying solely on time domain measurements, the results suggest that the robustness to parameter changes is increased while the tracking accuracy is enhanced by 50\% or more when 512 particles are used.
\end{abstract}

\section{Introduction} \label{S:introduction}

In the past years, dense wireless sensor network deployments have been exploited in novel sensing applications such as device-free localization (DFL) \cite{patwari2010} and non-invasive breathing monitoring \cite{patwari2011b}, and used by residential monitoring \cite{kaltiokallio2012b}, ambient assisted living \cite{bocca2013} and military \cite{bjorkbom2013} systems. The aforementioned narrowband systems use measurements of the radio to extract information about people in the surrounding environment and therefore, such networks are referred to as RF sensor networks \cite{patwari2010}. These systems typically use the received signal strength (RSS) measurements of the radio in order to quantify changes in the propagation channel. 

DFL systems aim at localizing and tracking people in a monitored area using the RSS of low-cost and static transceivers. For this purpose, there are two model based approaches that are widely utilized: an imaging approach referred to as \emph{radio tomographic imaging} (RTI) \cite{patwari2008}, and a Bayesian inference approach which is typically solved using sequential Monte Carlo methods \cite{li2011}. In both cases, the measured RSS is related to a person's kinematic state, i.e., position and velocity using a spatial RSS model. However, the time domain RSS measurements are only affected by instantaneous location of the person and therefore, the kinematic system is not expected to be completely observable. This in turn limits the performance of DFL systems. One solution to address this problem and to improve the accuracy of RSS-based DFL is to include measurements that are related to the person's velocity. In this paper, we show that the peak frequency component of RSS is a function of person's position and velocity.     

A moving person alters excess path length of the reflected signal which causes destructive and constructive fading sequentially resulting that the RSS varies periodically \cite{kaltiokallio2014b,yigitler2014}. This observation necessarily implies that the power spectral density (PSD) of the RSS has impulsive spikes at frequencies defined by the rate at which the propagation medium is being altered. The frequency of the highest spike where the PSD achieves its maximum, which we refer to as \emph{frequency domain RSS measurement} in this paper, is of particular interest since it is related to the person's position and velocity. We utilize this finding to estimate the person's kinematic state from the time and frequency domain RSS measurements. In addition to augmenting the tracking algorithm with the new measurement, the frequency domain RSS measurements are not as easily altered by noise, measurement equipment induced effects such as carrier and symbol synchronization, or electrical properties of the person. Thus, the benefits of using the frequency domain RSS measurements are two-fold.

In this paper, a particle filter is implemented to estimate kinematic state of a person from the time and frequency domain RSS measurements. The tracking performance of the filter is evaluated using simulations and the development efforts are validated with experimental data. Compared to a system relying solely on time domain measurements, the results suggest that the robustness to parameter changes is increased while the tracking accuracy is enhanced by $50\%$ or more when $512$ particles are used.

The rest of the paper is organized as follows. In the following section, the related work is discussed. In Section \ref{S:model}, models for the time and frequency domain RSS measurements are presented and the particle filter used for tracking is introduced. In Section \ref{S:simulation_results}, the system performance is evaluated using simulations and in Section \ref{S:experimental_results}, the development efforts are validated using experimental data. Requirements and possibilities of velocity estimation are discussed in Section \ref{S:discussion} and conclusions are drawn in \ref{S:conclusions}.

\section{Related Work} \label{S:related_work}

In indoor environments, the RSS measurements are typically dictated by multipath propagation causing the RSS measurements to differ from the modeled and therefore, it has a negative impact on localization accuracy \cite{kaltiokallio2014c}. To address this problem, one can attempt to derive a more profound spatial model that accounts for uncertainties in the environment and propagation mechanisms. Another possibility to improve the performance is to mitigate effects of the unknown factors by using special hardware features such as directional antennas \cite{wei2014}, channel diversity \cite{kaltiokallio2012a} or both \cite{kaltiokallio2014b}. A third possibility is to utilize robust measurement metrics for localization purposes and in this paper, the frequency domain RSS measurements are used. This metric is not as easily altered by noise, measurement equipment induced effects such as carrier and symbol synchronization, or electrical properties of the person. To date, the robustness of spectral estimation has been successfully used in RF sensor networks to monitor small breathing-induced changes in the propagation channel \cite{patwari2011b,patwari2014,kaltiokallio2014a}. 

In RSS-based DFL, the discrete-time constant velocity model \cite[Ch. 6]{BarShalom2001} is widely utilized to track the position and velocity of the person \cite{guo2013, zhao2011, kaltiokallio2014b}. Other commonly used models to describe the target dynamics include the one-tap autoregressive Gaussian model where velocity is treated as noise \cite{li2011}, and the jump-state Markov model which is suitable for describing the dynamics of a maneuvering object \cite{Nannuru2013}. For tracking, the aforementioned works solely rely in the time domain RSS measurements and therefore, the kinematic system is not expected to be completely observable. In this work, we also use the discrete-time constant velocity model to describe the dynamics of the person. However, we incorporate the frequency domain measurements to the tracking algorithm to include measurements of the person's velocity. 
 
In addition to the contributions listed above, we use the following features to maximize performance of the system in this paper. Directional antennas are used to limit the region where human-induced temporal fading is observed as in \cite{wei2014,kaltiokallio2014b}. In addition, channel diversity is exploited to increase the probability that at least one of the frequency channels follows the spatial model \cite{kaltiokallio2012a} and to increase signal-to-noise ratio of the measurements. Furthermore, the three-state measurement model proposed in \cite{kaltiokallio2014b} is utilized so that both shadowing and reflection mechanism are accounted for. 

\section{Methodology} \label{S:model}

\subsection{Preliminaries}

A typical narrowband receiver samples and outputs the RSS every $T_s$ seconds, so that the measurement at time $k$ on frequency channel $c$ can be expressed as
\begin{equation}\label{eq:rss}
	\hat{r}(k;c) = P(c) + h(k;c) + \hat{\nu}_r(k;c),
\end{equation}
where $P(c)$ is a communication system dependent gain, $h(k;c)$ the channel gain and $\hat{\nu}_r(k;c)$ is zero-mean additive wideband noise. If the channel is stationary for the duration of interest, the statistical expectation $\mathrm{E}\{\hat{r}(k;c)\}= P(c) + \mathrm{E}\{h(k;c)\}$ of the measurements is equivalent to the time average. Under such conditions, the expected value reflects the site dependent and time-invariant channel characteristics. Thus, removing the mean yields a signal which does not depend on the underlying measurement setup and environment. Thereafter, time variations of the mean removed RSS, given by
\begin{equation}\label{eq:mean_removed_rss}
\begin{aligned}
\tilde{r}(k;c) &= \hat{r}(k;c) - \mathrm{E}\{\hat{r}(k;c)\} \\ 
	&= h(k;c) - \mathrm{E}\{h(k;c)\} + \hat{\nu}_r(k;c)
\end{aligned}
\end{equation}
reflects the possible changes in the propagation channel. 

In this paper, the channels are combined by averaging to increase the signal-to-noise ratio (SNR) of the measurements. The mean-removed and combined RSS measurements are denoted as
\begin{equation} \label{eq:pre_processed_rss}
    r(k) =  g(k) + \nu_r(k),
\end{equation}
where $g(k) = \sum_{c=1}^C h(k;c) - \mathrm{E}\{h(k;c)\}$, $\nu_r(k)$ is i.i.d. zero-mean additive wideband noise and $C$ is the number of channels. It is to be noted that averaging may not be the best method to combine the channels. This is because the measurements can differ considerably among the channels when using omni-directional antennas \cite{kaltiokallio2012a} or when the person is far away from the LoS resulting that phase of the reflected signals are considerably different \cite{yigitler2014}. Averaging is justifiable for the development efforts of this paper because directional antennas are used and the person is tracked in the close vicinity of the LoS. In such scenarios, the RSS measurements are expected to correlate highly among each other on the different channels.

\subsection{RSS in Time Domain}

Kaltiokallio \emph{et. al} have recently presented that the temporal RSS changes are caused by: i) electronic noise, when a person is not present in the monitored area; ii) reflection, when a person is moving in the close proximity of the LoS; iii) shadowing, when a person is obstructing the LoS \cite{kaltiokallio2014b}. Based on the findings, a three-state temporal RSS model was presented where $g(k)$ in Eq.~\eqref{eq:pre_processed_rss} is represented with respect to the state of the propagation medium as
\begin{equation}\label{eq:three_state_model}
    g(k) =  \begin{cases}
	g_{1}(\bm{p}) \quad \text{ (\emph{$s_1$ = non-fading})},\\
	g_{2}(\bm{p}) \quad \text{ (\emph{$s_2$ = reflection})},\\
	g_{3}(\bm{p}) \quad \text{ (\emph{$s_3$ = shadowing})},\\
               \end{cases}
\end{equation}
where $\bm{p} = \left[p_x \quad p_y\right]^T$ represents coordinates' of the person and $s_i$ state of the propagation channel. 

In non-fading state, the propagation channel is stationary and $\nu_r(k)$ dictates $r(k)$ because a single realization of the fading process is observed. Since the mean is removed, we have $g_{1}(\bm{p}) = 0$. In reflection state, $r(k)$ is dominated by single-bounce reflections and $g_{2}(\bm{p})$ is given by \cite{kaltiokallio2014b}
\begin{equation}\label{eq:reflection_model}
    g_{2}(\bm{p}) = 10 \log_{10} \left(\Psi^2 + 2 \Psi\cos\phi_r(k)  +1 \right),
\end{equation}
where $\phi_r(k)$ is phase of the reflected signal and $0<\Psi<1$ describes the relation between the LoS component and the reflected component. In shadowing state, shadowing dominates $r(k)$ and $g_{3}(\bm{p})$ can be represented by a \emph{line integral} traversing through the person along a straight line from TX to RX. If the person is modeled using an ellipse with uniform electrical properties, the total attenuation along the LoS is \cite[Ch. 3]{Kak1988}
\begin{equation}\label{eq:projection}
    g_{3}(\bm{p})  = \begin{cases}
                       \frac{2{\rho}AB}{a^2(\theta)}\sqrt{a^2(\theta) - (p_x)^2} & \text{if } \lvert p_x \rvert \leq a(\theta)  \\
                       0 & \text{otherwise}
               \end{cases}
\end{equation}
where $\rho$ is the attenuation factor, $A$ and $B$ are the semi-minor and semi-major axis of the ellipse, $\theta$ is rotation of the person with respect to the transceivers and $a^2(\theta) = A^2\cos^2(\theta) + B^2\sin^2(\theta)$. For further details about the spatial models, the readers are referred to \cite{kaltiokallio2014b} where the models were originally derived.

\subsection{RSS in Frequency Domain}

\subsubsection{Modeling the Propagation Channel's Change Rate}

In general, path length of different propagation paths changes with time. This change can be due to movement of the TX, RX, dynamic objects in the environment, or any combination of the aforementioned \cite[Ch. 5]{Molisch2010}. As an outcome, phases of the received signals vary periodically at the RX causing time-varying changes in the RSS referred to as \emph{small-scale fading}. When the transceivers are stationary, this phenomenon is effective when the person moves in close vicinity but not on the LoS. In other cases, either the RSS is composed of random variations or it has a strong and almost constant decrease due to shadowing. Therefore, the frequency content of the RSS is informative only in reflection state, i.e., when
$g(k) = g_{2}(\bm{p}) $ in Eq.~\eqref{eq:three_state_model}.

In the single-bounce reflection case, excess path length of the reflected signal $\Delta_r(t)$ determines the phase as $\phi_r(t) = 2\pi\Delta_r(t)/\lambda$ where $\lambda$ is wave length of the carrier frequency. Since the RSS of a single-bounce reflection is a periodic function, in \cite{yigitler2014} it has been shown that its trigonometric Fourier series representation is 
\begin{equation}\label{eq:fourier_series}
    g_2(\Delta_r(t)) = -2\hat{e} \sum_{i=1}^{\infty}{a_i \cos \left( 2 \pi i \frac{\Delta_r(t)}{\lambda}\right)},
\end{equation}
in which $a_i \triangleq (-\Psi)^i/i$ and $\hat{e} = 10 \log_{10} (e)$, where $e$ is the base of the natural logarithm; $\lambda = f_c/c_0$, where $f_c$ is the carrier frequency and $c_0$ is the speed of light in free space. 

When the person moves with non-zero velocity at time $t_0$, $\Delta_r(t)$ is a continuous function of time so that its Taylor series expansion around $t_0$ is given by
\begin{equation}\label{eq:taylor_series}
    \Delta_r(t) = \sum_{i=0}^{\infty}{\frac{\Delta_r^{(i)}(t_0)}{i!}\left(t-t_0\right)^i},
\end{equation}
where superscript $\left(i\right)$ denotes the order of differentiation. Consequently, for constant velocity movements we have
\begin{equation}\label{eq:constant_velocity_taylor_series}
    \Delta_r(t) = \Delta_r(t_0) + \dot{\Delta}_r(t - t_0) + \mathcal{O}\left((t-t_0)^2 \right),
\end{equation}
where $\dot{\Delta}_r = \frac{d}{dt}\Delta_r(t) \vert _{t=t_0}$ and error $\mathcal{O}\left((t-t_0)^2 \right)$ is a function of $\ddot{\Delta}_r = \frac{d^2}{dt^2}\Delta_r(t) \vert _{t=t_0}$. In case the second and higher order terms can be ignored, plugging Eq.~\eqref{eq:constant_velocity_taylor_series} into Eq.~\eqref{eq:fourier_series} yields
\begin{equation}\label{eq:constant_velocity_approximation}
	g_2(\Delta_r(t)) \approx -2\hat{e} \sum_{i=1}^{\infty}{a_i \cos \left( 2 \pi \frac{i}{\lambda} \dot{\Delta}_r t  + \varphi_i \right)}.
\end{equation}
The constant term $\varphi_i$ is the phase of the cosine functions and it has no impact on the amplitude spectrum of $g(k) = g_2(\Delta_r(t))$. Consequently, the spectrum has peaks around $\frac{i}{\lambda}\dot{\Delta}_r$ and locally the spectrum has the form of a Dirac-delta convolved with the spectrum of $a_i$.

Since $\Psi<1$, we have $a_1 > a_i$ for $i=2,3,\cdots$. This implies that the peak of the frequency spectrum of $g(k)$ is around the instantaneous frequency of the first cosine function in Eq.~\eqref{eq:constant_velocity_approximation}, that is
\begin{equation}\label{eq:instantaneous_frequency}
    G(t) = \frac{1}{\lambda}\frac{d}{d t}\Delta_r (t),
\end{equation}
where excess path length of the reflected signal is defined as
\begin{equation}\label{eq:excess_path_length}
    \Delta_r (t) = \lVert \bm{p}_r - \bm{p}_{rx} \rVert + \lVert \bm{p}_r - \bm{p}_{tx} \rVert - \lVert \bm{p}_{rx} - \bm{p}_{tx} \rVert,
\end{equation}
and $\bm{p}_r$, $\bm{p}_{rx}$ and $\bm{p}_{tx}$ are coordinates of the reflection point, receiver and transmitter in corresponding order. When the transceivers are stationary, only the velocity of the reflector, $\boldsymbol{v}$, effects $G(t)$ in Eq.~\eqref{eq:instantaneous_frequency}, i.e.,
\begin{equation}\label{eq:rate_of_change}
    \frac{d}{d t}\Delta_r(t) =
    \frac{1}{\lVert \bm{p}_r - \bm{p}_{rx} \rVert} \begin{bmatrix} p_{x,r} - p_{x,rx} \\ p_{y,r} - p_{y,rx} \end{bmatrix}^T\bm{v} + 
    \frac{1}{\lVert \bm{p}_r - \bm{p}_{tx} \rVert} \begin{bmatrix} p_{x,r} - p_{x,tx} \\ p_{y,r} - p_{y,tx} \end{bmatrix}^T\bm{v}.
\end{equation}
In Eq.~\eqref{eq:rate_of_change} the chain rule of differentiation has been used to derive the equation from Eq.~\eqref{eq:excess_path_length}. Now, Eq.~\eqref{eq:rate_of_change} can be compactly expressed as 
\begin{equation}\label{eq:rate_of_change_compressed}
\begin{aligned}
    \frac{d}{d t}\Delta_r(t) &= \left( \tilde{\bm{p}}_{rx} + \tilde{\bm{p}}_{tx}  \right)^T \bm{v}, \quad \text{where}\\
\tilde{\bm{p}}_{rx} &\triangleq \frac{\bm{p}_r - \bm{p}_{rx}}{\lVert \bm{p}_r - \bm{p}_{rx} \rVert}, \\
\tilde{\bm{p}}_{tx} &\triangleq \frac{\bm{p}_r - \bm{p}_{tx}}{\lVert \bm{p}_r - \bm{p}_{tx} \rVert},
\end{aligned}
\end{equation}
in which $\tilde{\bm{p}}_{rx}$ and $\tilde{\bm{p}}_{tx}$ have unity norm and the derivative only exists when $\bm{p}_r \neq \bm{p}_{rx}$ and  $\bm{p}_r \neq \bm{p}_{tx}$. Since $a_1 = -\Psi$  has a very strong DC component, convolving its spectrum with the Dirac-delta function does not change the frequency of the peak. Therefore, the strongest frequency component of $g(k)$ is a function of both velocity and position of the reflector, and it does not depend on $\Psi$. As a consequence, $G(k) = G(k T_s)$ does not depend on the irregularities of the communication system, large scale losses of the environment, and electrical properties of the reflector. 

\subsubsection{Measuring the Propagation Channel's Change Rate}

In order to quantify the propagation channel's change rate in Eq.~\eqref{eq:rate_of_change_compressed} we use a sequence of $\{r(k)\}_{k=1}^{N_f}$ to calculate the power spectral density (PSD) of the measurements, and find the frequency of its peak, i.e., 
\begin{equation}\label{eq:modeled_frequency}
    R(k) = \argmax_f \left\{\left \lvert \sum_{k=1}^{N_f}r(k) \cdot e^{-j 2 \pi f k T_s} \right \rvert ^2 \right\},
\end{equation}
where $f$ is the frequency, $T_s$ is the sampling interval and $N_f$ is the number of samples. For the reasons discussed in the preceding section, now we have
\begin{equation}\label{eq:rss_frequency}
    R(k) = \frac{1}{\lambda}\left( \tilde{\bm{p}}_{rx} + \tilde{\bm{p}}_{tx}  \right)^T \bm{v} + \nu_f(k),
\end{equation}
where $\nu_f(k)$ is zero-mean additive measurement noise. 

The noise $\nu_f(k)$ depends on the statistical properties of the frequency estimate. The intensity of $\nu_f(k)$ is dictated by variance of $\nu_r(k)$ in Eq.~\eqref{eq:pre_processed_rss}, sampling rate, number of DFT points, higher order harmonics in Eq.~\eqref{eq:fourier_series}, and higher order terms in Eq.~\eqref{eq:constant_velocity_taylor_series}. Thus, for signal-to-noise ratio values high enough not to cause thresholding effect \cite{Rife1974} and when the higher order terms in Eq.~\eqref{eq:constant_velocity_taylor_series} can be ignored, $\nu_f(k)$ may be assumed independent of  $\nu_r(k)$. Furthermore, if the number of DFT points is high enough and under the aforementioned conditions, $\nu_f(k)$ can be taken as a normal variate. Consequently, we assume that $\nu_f(k)$ is a zero-mean Gaussian. 

It is to be noted that since $R(k)$ is evaluated over a time window of length $N_f$, it measures the average rate of change over this time window. To model this appropriately, also $G(k) = G(kT_s)$ has to be evaluated in the \emph{average position} over time window $N_f$. For constant velocity $\bm{v}$, this is $G(k-\frac{N_f}{2})$.

\subsection{Localization and Tracking}

In RSS-based DFL, we are interested in the location and trajectory of the person in an inertial frame of reference. This problem can be formulated using the widely utilized discrete-time constant velocity model \cite[Ch. 6]{BarShalom2001} which represents the evolution of state 
$$
\bm{x}(k) = [p_x(k) \quad v_x(k) \quad p_y(k) \quad v_y(k)]^T
$$
as 
\begin{equation} \label{eq:state_model}
	\bm{x}(k) = \bm{A} \bm{x}(k-1) + \bm{B} \bm{u}(k-1),
\end{equation}
where $\bm{u}$ is zero-mean i.i.d Gaussian process noise and 
$$
\bm{A} = \left[\begin{matrix} 1&T_s&0&0 \\ 0&1&0&0 \\ 0&0&1&T_s \\ 0&0&0&1 \end{matrix}\right] \negthickspace\text{,}\quad
\bm{B} = \left[\begin{matrix} \frac{1}{2}T_s^2&0 \\ T_s&0 \\ 0&\frac{1}{2}T_s^2 \\ 0&T_s\end{matrix}\right] \negthickspace\text{.}\quad
$$
The objective of tracking is to estimate $\bm{x}(k)$ using the sequence of noisy measurements
\begin{equation} \label{eq:measurement_model}
\begin{aligned}
	\bm{z}(k) &= \bm{h}(k) + \bm{\nu}(k), \\
	\bm{h}(k) &= [g(\bm{x}(k)) \quad G(\bm{x}(k))]^T,
\end{aligned}
\end{equation}
where $\bm{\nu}(k)$ is i.i.d measurement noise. We approximate $\bm{\nu}(k)$ as zero-mean multivariate Gaussian, i.e.,  $\bm{\nu}(k) \sim \mathcal{N}_2(\bm{0},\bm{\Sigma})$. In this case, the multivariate normal distribution with two degrees of freedom can be used to characterize the noise as
\begin{equation} \label{eq:residual_density}
f(\bm{\nu}(k) \vert \bm{0},\bm{\Sigma}) = \frac{1}{\sqrt{(2\pi)^2\lvert \bm{\Sigma} \rvert }} \text{exp} \left( -\frac{1}{2}\bm{\nu}(k)^T\bm{\Sigma}^{-1}\bm{\nu}(k) \right),
\end{equation}
where $\bm{\Sigma}$ is the covariance matrix, $\lvert \cdot \rvert$ denotes the determinant and $(\cdot)^T$ the transpose.

\subsection{Application}

In this work, we use a hidden Markov model (HMM) to estimate temporal state of the propagation channel as in \cite{kaltiokallio2014b}. The temporal state estimates are used to identify the time instances when the state changes from reflection to shadowing so that the tracking filter can be initialized when the person is in the close vicinity of the LoS. Correspondingly, the estimates are used to identify when the state changes from reflection to non-fading state so that tracking can be ended when the person is far away from the LoS. The pseudo-code of the overall application is presented in Algorithm $1$.

\begin{center}
\renewcommand{\arraystretch}{1.2}
  \begin{tabular}{ l }
    \hline\hline
	\bf{Algorithm 1: } \it{Application} \\
    \hline
	\it{At time k, use HMM to estimate link state $\hat{s}(k)$} \\
	\bf{if} \it{particle filter is initialized} \bf{do} \\
		\quad \bf{if} $\hat{s}(k) = s_1 \; \text{ \it{ for all links}}$ \bf{do} \it{stop tracking} \\
		\quad \bf{else} \it{estimate kinematic state} $\bm{\hat{x}}_k$ \it{using} \bf{Algorithm 2}\\
	\bf{elseif} $\hat{s}(k) = s_3 \text{ \it{ for any link}}$ \bf{do} \it{initialize particle filter} \\
	\hline
  \end{tabular}
\end{center}

We exploit a particle filter for tracking the evolution of state $\bm{x}(k)$. Particle filters are based on point mass representation of probability densities and they are especially suitable for non-linear/non-Gaussian problems where optimal algorithms such as the Kalman filter fail \cite{Arulampalam2002}. Moreover, they have been successfully used in DFL applications to track the movements of a person e.g. in \cite{li2011,kaltiokallio2014b,Wilson2012,Nannuru2013}. The implemented particle filter is summarized in Algorithm $2$.

\begin{center}
\renewcommand{\arraystretch}{1.3}
  \begin{tabular}{ l }
    \hline\hline 
	\bf{Algorithm 2: } \it{Particle Filter} \\
    \hline

\bf{if} \it{particle filter is initialized} \bf{do} \\

(1) \textit{Predict} state  $\bm{x}_i(k)$ using Eq.~\eqref{eq:state_model} for particle $i$ \\
(2) \textit{Measure} $\bm{z}(k)$ and calculate $\bm{h}_i(k)$ using \eqref{eq:three_state_model} and \eqref{eq:instantaneous_frequency}\\
(3) \textit{Update} weight $w_i(k) \propto f(\bm{\nu}_i(k) \vert \bm{0},\bm{\Sigma})$ using Eq.~\eqref{eq:residual_density} \\ 
(4) \textit{Normalize} $\tilde w_i(k) = \frac{w_i(k)}{\sum_{i=1}^{N} w_i(k)} \rightarrow \sum_{i=1}^{N} \tilde{w}_i(k) = 1$  \\
(5) \textit{Resample} particles $\lbrace \bm{x}_i(k), \tilde{w}_i(k) \rbrace \rightarrow  \lbrace \frac{1}{N}, \bm{x}_i(k) \rbrace$ \\ 
(6) \textit{Estimate} $\bm{\hat{x}}(k) = \frac{1}{N}\sum_{i=1}^{N} \bm{x}_i(k)$ as mean of particles \\

\bf{else} \it{initialize $\bm{x}(k)$} \\

\textit{Initialize} $\bm{\check{p}}_i$ as in \cite{kaltiokallio2014a} and $\bm{\check{v}}_i = \left[\check{v}\cos(\check{\theta}) \quad \check{v}\sin(\check{\theta}) \right]^T$, \\ 
where $\check{v} = \lVert \bm{v} \rVert \sim \mathcal{U}(0,2)$ and $\check{\theta} \sim \mathcal{U}(\theta - \frac{\pi}{4},\theta + \frac{\pi}{4})$ \\
\textit{Compute} steps (2)-(6) to estimate $\bm{\hat{v}}(k)$ \\
\textit{Substitute} $\bm{x}(k) = [\check{p}_x \quad \hat{v}_x(k) \quad \check{p}_y \quad \hat{v}_y(k)]^T $ \\

	\hline
  \end{tabular}
\end{center}

\begin{figure}[!t]
\begin{centering}
\includegraphics[]{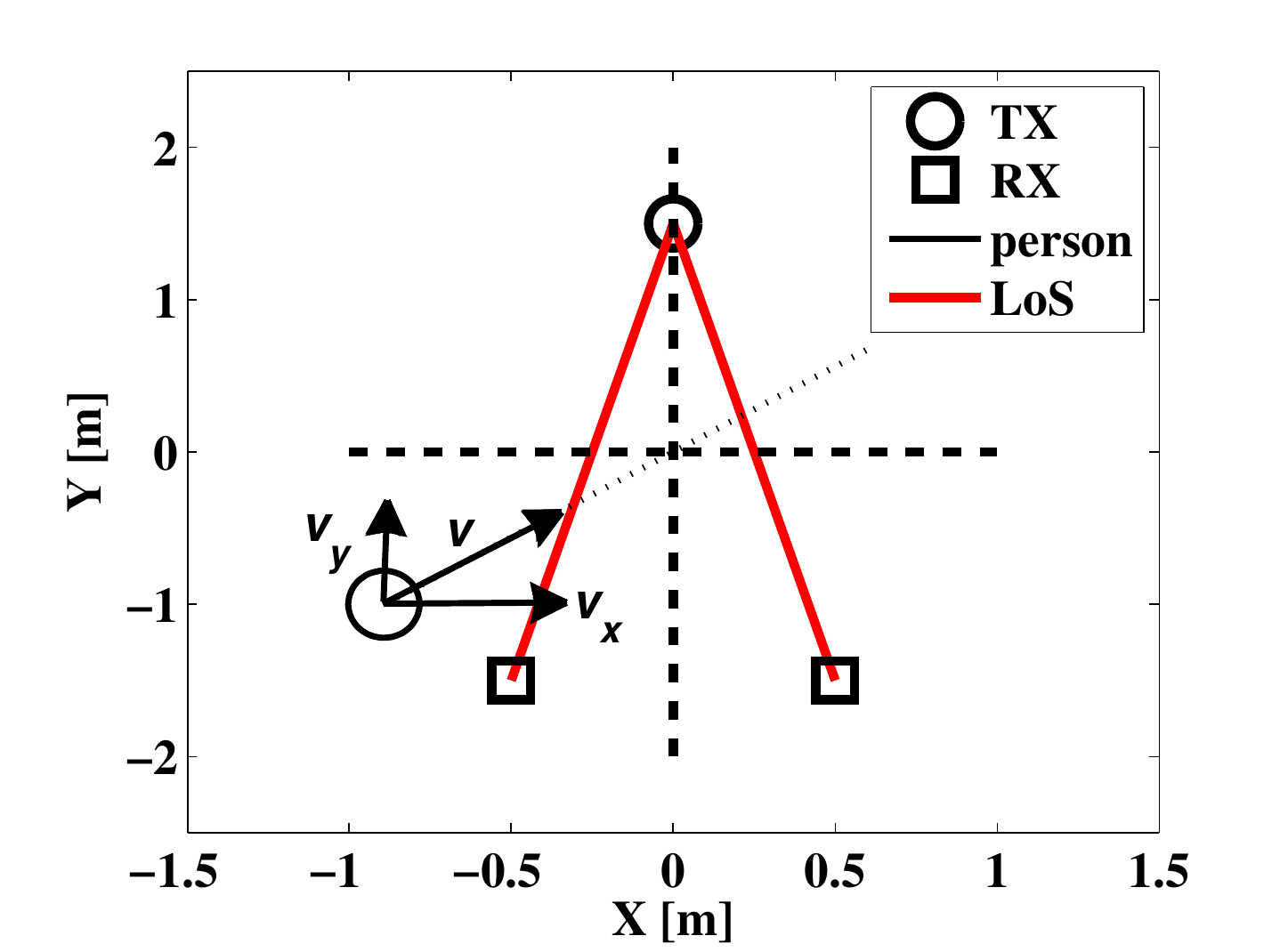}
\caption{Experimental setup} 
\label{fig:experimental_setup}
\end{centering}
\end{figure}

\section{Simulation Results} \label{S:simulation_results}

\subsection{Simulation Description}

\begin{figure*}[t]
\begin{centering}
\begin{tabular}{ccc}

\subfloat[With $R(k)$]{\includegraphics[width=\columnwidth*30/100]{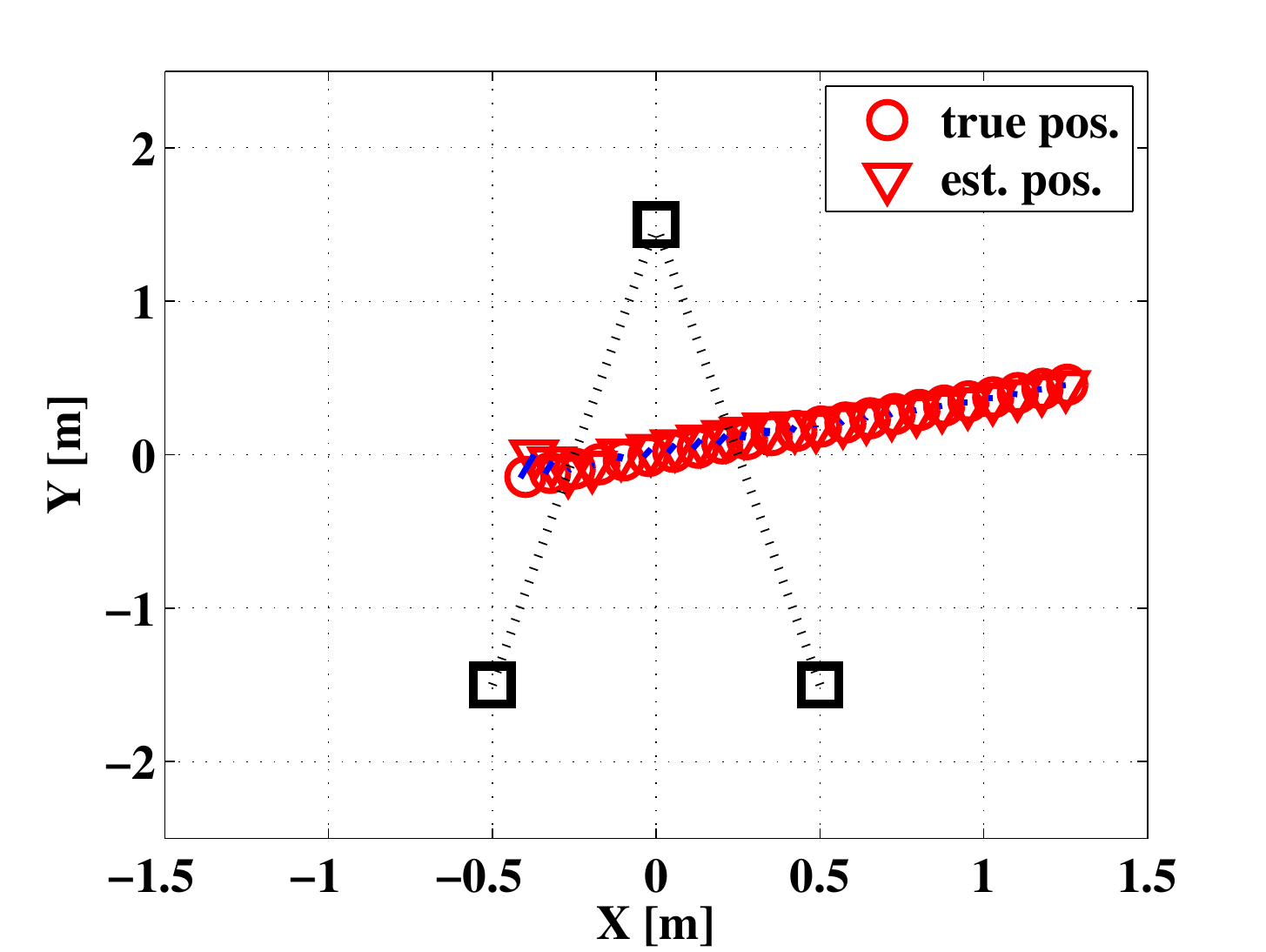}\label{fig:trajectory1}}
\subfloat[Without $R(k)$]{\includegraphics[width=\columnwidth*30/100]{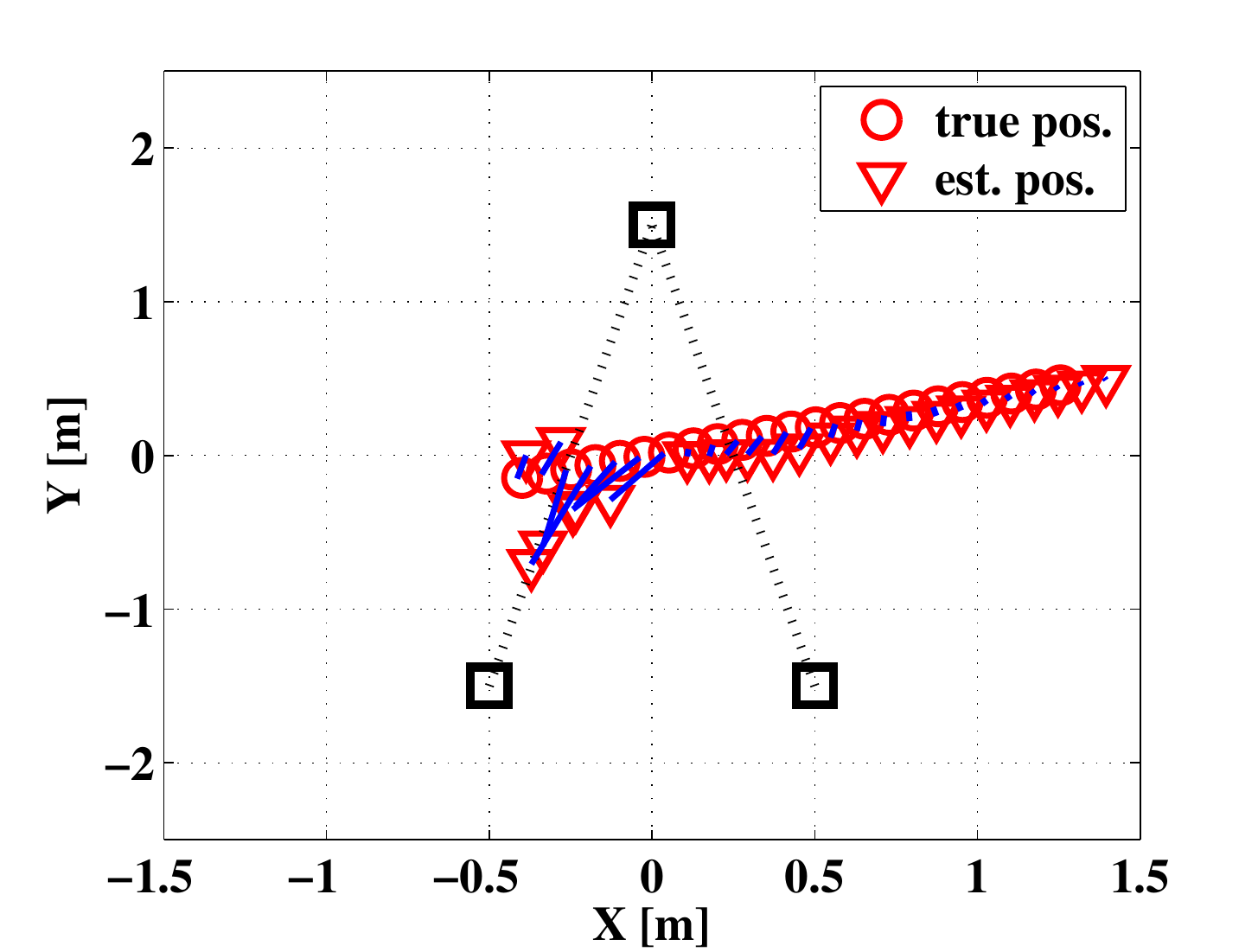}\label{fig:trajectory2}}
\subfloat[$\bar{\varepsilon}_\%$ as function of $\theta$ and $\nu_r$]{\includegraphics[width=\columnwidth*30/100]{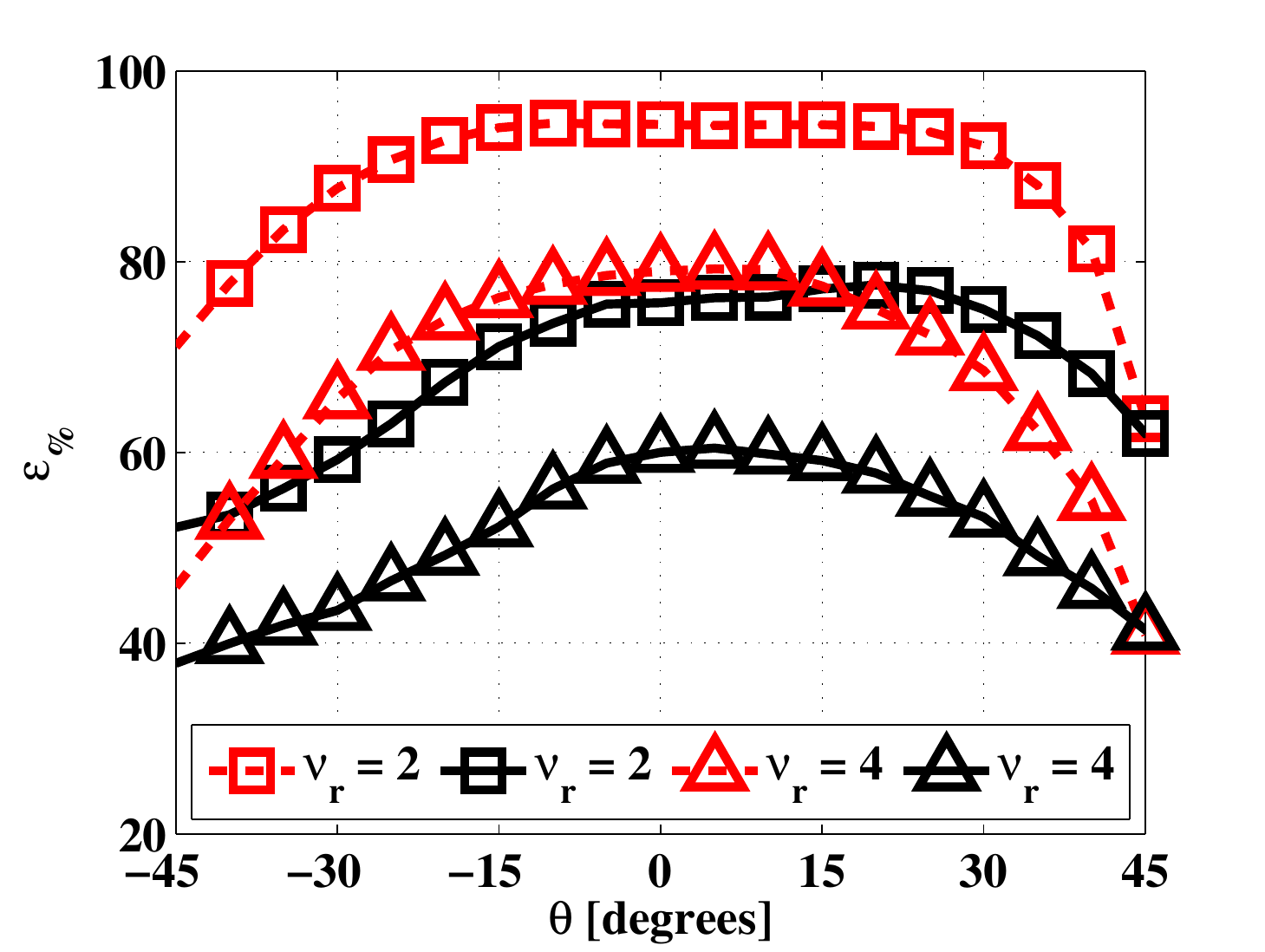}\label{fig:theta_effect}}

\end{tabular}
\caption{Example tracking performance using the frequency content of the measurements in (a) and without them in (b) when the target's heading is $\theta = 20^\circ$. In (c), $\bar{\varepsilon}_\%$ with (red) and without (black) the frequency domain measurements as a function of $\theta$ and with various measurement noise $\nu_r$ values.} 
\label{fig:simulation_results}
\end{centering}
\end{figure*}

Incorporating the frequency domain measurements in the tracking filter is evaluated in this section. For the purpose, the effect of measurement noise $\hat{\nu}_r(k;c)$ and target's heading $\theta = \tan^{-1}\left(v_y/v_x\right)$ is studied and the results are compared with respect to a system that does not utilize the frequency content of the measurements. In the simulations, the human target is moving with constant speed $v = \lVert \bm{v} \rVert = 0.5 \text{ m/s}$ and the trajectory bisects the origin with various angles, i.e., $\theta= \left[-\frac{\pi}{4}, \frac{\pi}{4} \right]$. The object is tracked using one transmitter and two receivers as depicted in Fig.~\ref{fig:experimental_setup}. The TX is set to transmit packets over each of the $16$ frequency channels defined by the IEEE 802.15.4 standard \cite{802_15_4} so that sampling rate of each frequency channel is $T_s = 32 \text{ ms}$. 

Parameters of the particle filter are: $N = 512$, $\bm{\Sigma} = [2.0 \quad 0; \quad 0 \quad 1.5]$ and $\bm{u} \sim \mathcal{N} (0, \sigma^2 \bm{I})$, where $\sigma = 0.4$. Window length to calculate the frequency content of the measurements in Eq.~\eqref{eq:modeled_frequency} is $N_f = 20$. Parameters of the spatial models used in Eqs.~\eqref{eq:reflection_model} and \eqref{eq:projection} are the same as reported in \cite{kaltiokallio2014b}. When the particle filter is initialized, the target's speed is set to $\check{v}= \lVert \bm{v} \rVert \sim \mathcal{U}(0,1)$ whereas the direction of movement to $\check{\theta} \sim \mathcal{U}(\theta - \frac{\pi}{4},\theta + \frac{\pi}{4})$, where $\theta$ depicts the true heading. 

\subsection{Evaluation Metrics}

The tracking accuracy is evaluated using mean absolute error (MAE) of the coordinate estimates 
\begin{equation} \label{eq:coordinate_error}
	\bar{\varepsilon}_x = \frac{1}{K} \displaystyle\sum\nolimits_{k=1}^K \left\lvert p_x(k) - \hat{p}_x(k) \right\rvert, 
\end{equation}
where $p_x$ depicts the true x-coordinate, $\hat{p}_x$ the estimate and $K$ the total number of estimates. The MAE for the y-coordinate, $\bar{\varepsilon}_y$, is calculated correspondingly. In addition, we use the percentage of particles within the modeled human ellipse as an evaluation criterion. The ratio is defined as 
\begin{equation} \label{eq:ratio}
	\bar{\varepsilon}_\% = \frac{ \sum_{k=1}^K N_{i \in \mathcal{A}}(k)}{K \cdot N} \cdot 100 \%,
\end{equation}
where $N_{i \in \mathcal{A}}(k)$ is the number of particles within area $\mathcal{A}$ spanned by the human ellipse model. The ratio indicates whether the posterior density has converged to the correct one or not. If the particles have converged to the correct density, it is expected that majority of the particles are located inside the human ellipse.

\subsection{Numerical Results}

In Figs.~\ref{fig:trajectory1} and \ref{fig:trajectory2}, example tracking performance of the system with and without the frequency domain measurements when the target's speed is $v = 0.5 \text{ m/s}$ and heading $\theta = 20^\circ$. On average, the system that uses both the time and frequency content of the RSS achieves higher tracking accuracy and converges faster to the correct trajectory. However, the improvement in tracking accuracy diminishes when $|\theta|$ increases as shown in Fig.~\ref{fig:theta_effect} and which we elaborate in the following.

The average variation of $\bar{\varepsilon}_\%$ with the direction of movement $\theta$ is shown in Fig.~\ref{fig:theta_effect}. The result suggests that as $|\theta|$ increases, the impact of $R(k)$ on tracking performance decreases. This can be well described by considering the scalar product in Eq.~\eqref{eq:rss_frequency}. First note that the unity norm vector sum, $ \tilde{\bm{p}}_{rx} + \tilde{\bm{p}}_{tx} $, is inward pointing normal vector of the Fresnel ellipse passing the point $\boldsymbol{p}_r$. Second, when the tracking is initialized, the person is close to the LoS so that this normal vector is approximately parallel to the semi-minor axis of the Fresnel ellipse for all considered $\theta$ values. This implies that the component of $\boldsymbol{v}$ perpendicular to the inward pointing normal increases with $|\theta|$. Now, one can verify that due to the last observation, the second order term of the Taylor series expansion in Eq.~\eqref{eq:constant_velocity_taylor_series} increases, causing larger and larger errors in the first order approximation. Therefore, the variance of $R(k)$ also grows, which in turn diminishes the improvement due to inclusion of $R(k)$.

\section{Experimental Results} \label{S:experimental_results}

\subsection{Experiment Description}

The presented system is experimentally validated using one transmitter and two receivers deployed at the opposite walls of a corridor resembling the deployment illustrated in Fig.~\ref{fig:experimental_setup}. Overall, the system is tested in three different corridors each having different width, i.e., $2.0 \text{, } 3.0 \text{ and } 3.5 \text{ m}$ and they are labeled as \emph{Experiment} $1$, $2$ and $3$ in respective order. The receivers are placed a meter away from each other. In each corridor, controlled experiments are conducted with a person moving at a known velocity ($v_x = \pm 0.5 \text{ m/s} \text{ and } v_y = 0 \text{ m/s}$) along a path defined \emph{a priori} of deployment. The path intersects the LoS of the links multiple times at various y-coordinates. 

The nodes are equipped with Texas Instruments CC2431 IEEE 802.15.4 PHY/MAC compliant $2.4 \text{ GHz}$ transceivers where the transceiver's micro-controller units run a communication software and a modified version of the FreeRTOS micro-kernel operating system. In the experiments, directional antennas are used to assure that the RX synchronizes to the LoS component and that the changes in RSS only reflect variations in the close vicinity of the LoS. The used directional antennas provide a $8$ dBi gain and a horizontal beam width of $75^{\circ}$ \cite{lcom}. A detailed description of the node hardware can be found at \cite{yigitler2014a}. 

The TX is programmed to transmit packets over each of the $16$ frequency channels defined by the IEEE 802.15.4 standard \cite{802_15_4} and after each transmission, the TX changes the frequency channel of communication in sequential order. The receivers are programmed to receive the packets and to store the data to a SD card for offline analysis. On average, the reception interval between two consecutive receptions is $2 \text{ ms}$ resulting that the sampling interval $T_s$ of each frequency channel is $32 \text{ ms}$. The communication protocol is the single transmitter version of the communication protocol described in \cite{yigitler2013}.

\subsection{Quantitative Evaluation}

\begin{figure*}[t]
\begin{centering}
\begin{tabular}{ccc}

\subfloat[True location]{\includegraphics[width=\columnwidth*30/100]{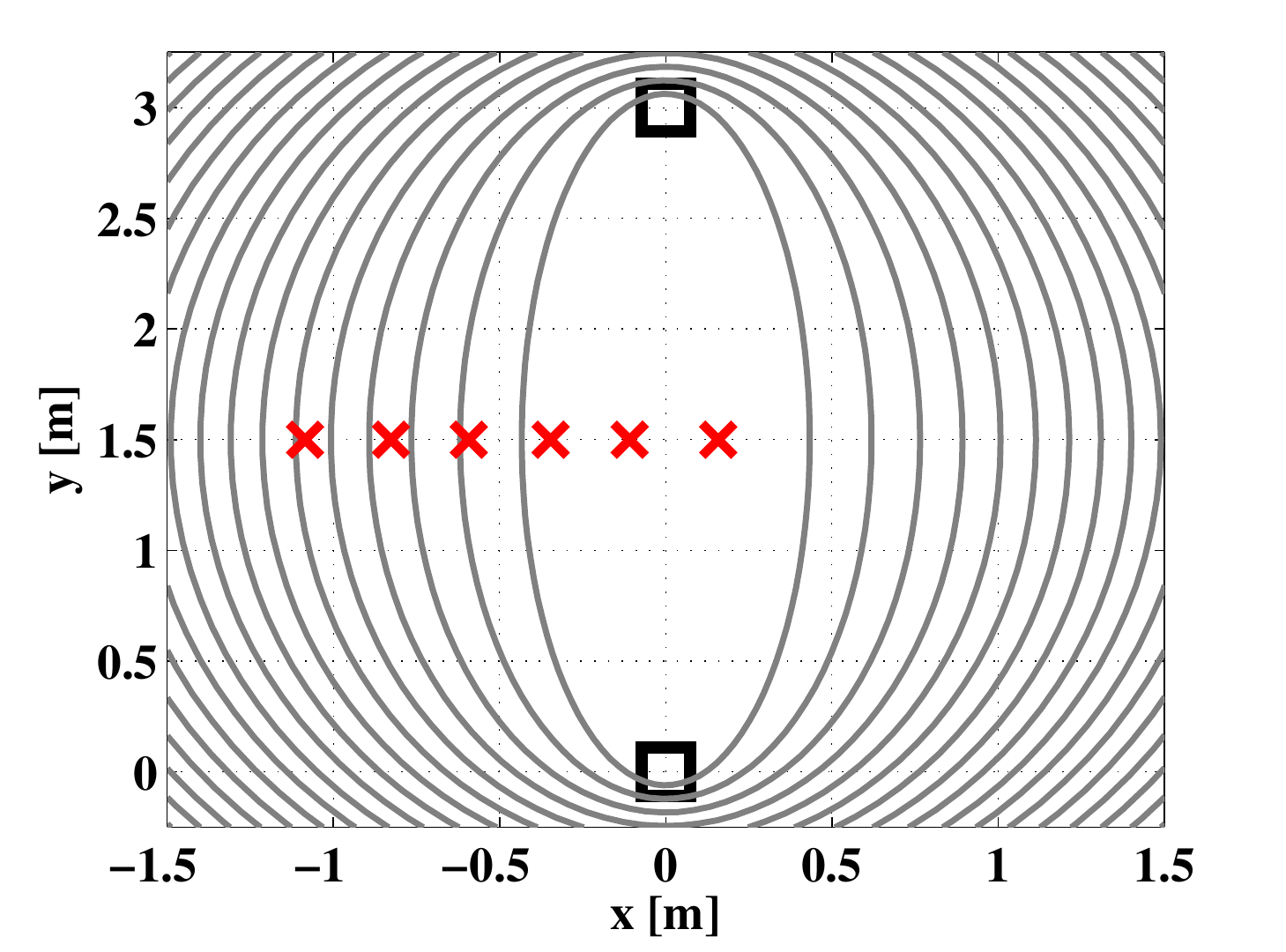}\label{fig:freznel_zones}}
\subfloat[Time domain, $r(k)$ vs. $g(k)$]{\includegraphics[width=\columnwidth*30/100]{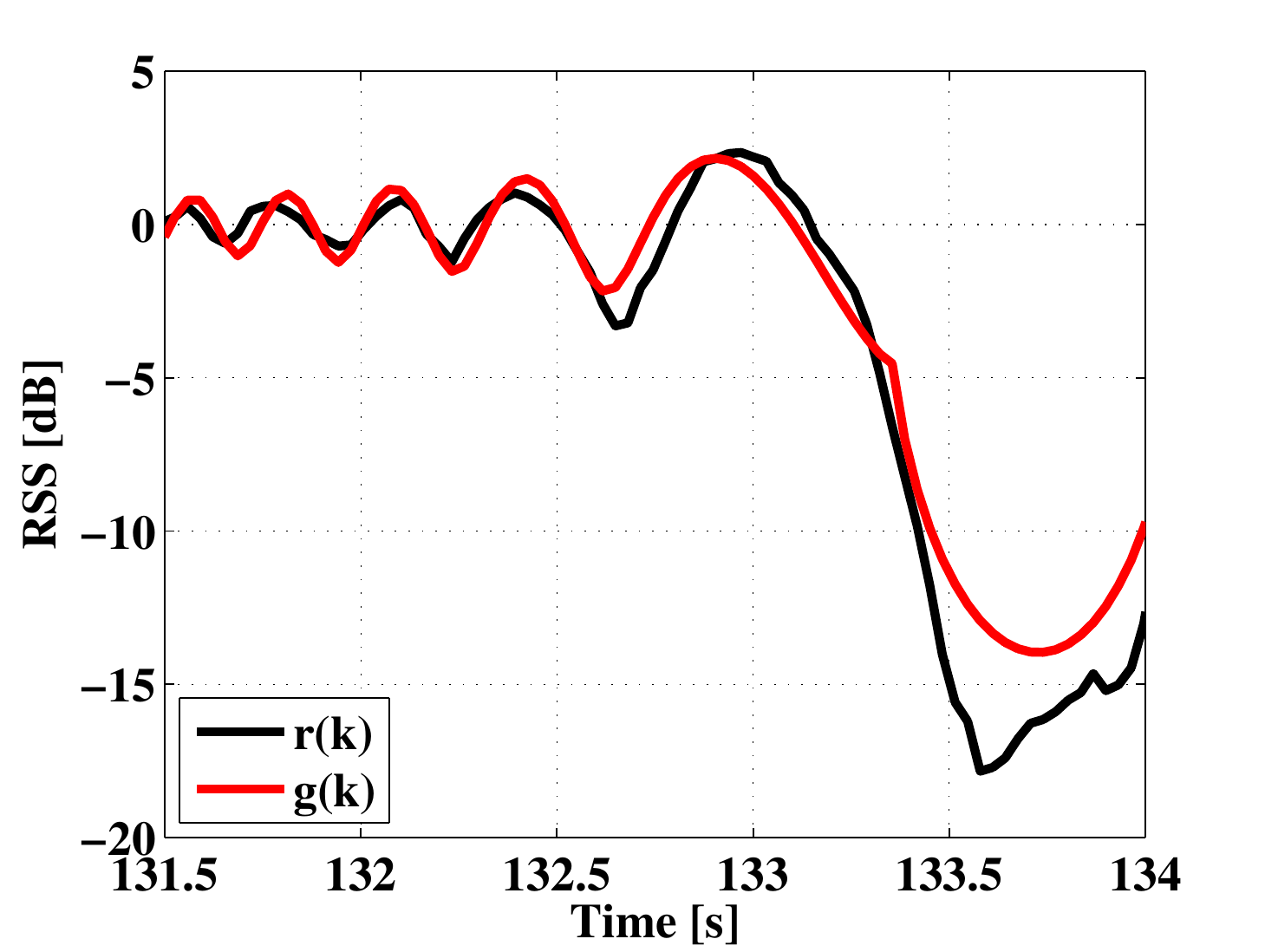}\label{fig:time_domain}}
\subfloat[Frequency domain, $R(k)$ vs. $G(k)$]{\includegraphics[width=\columnwidth*30/100]{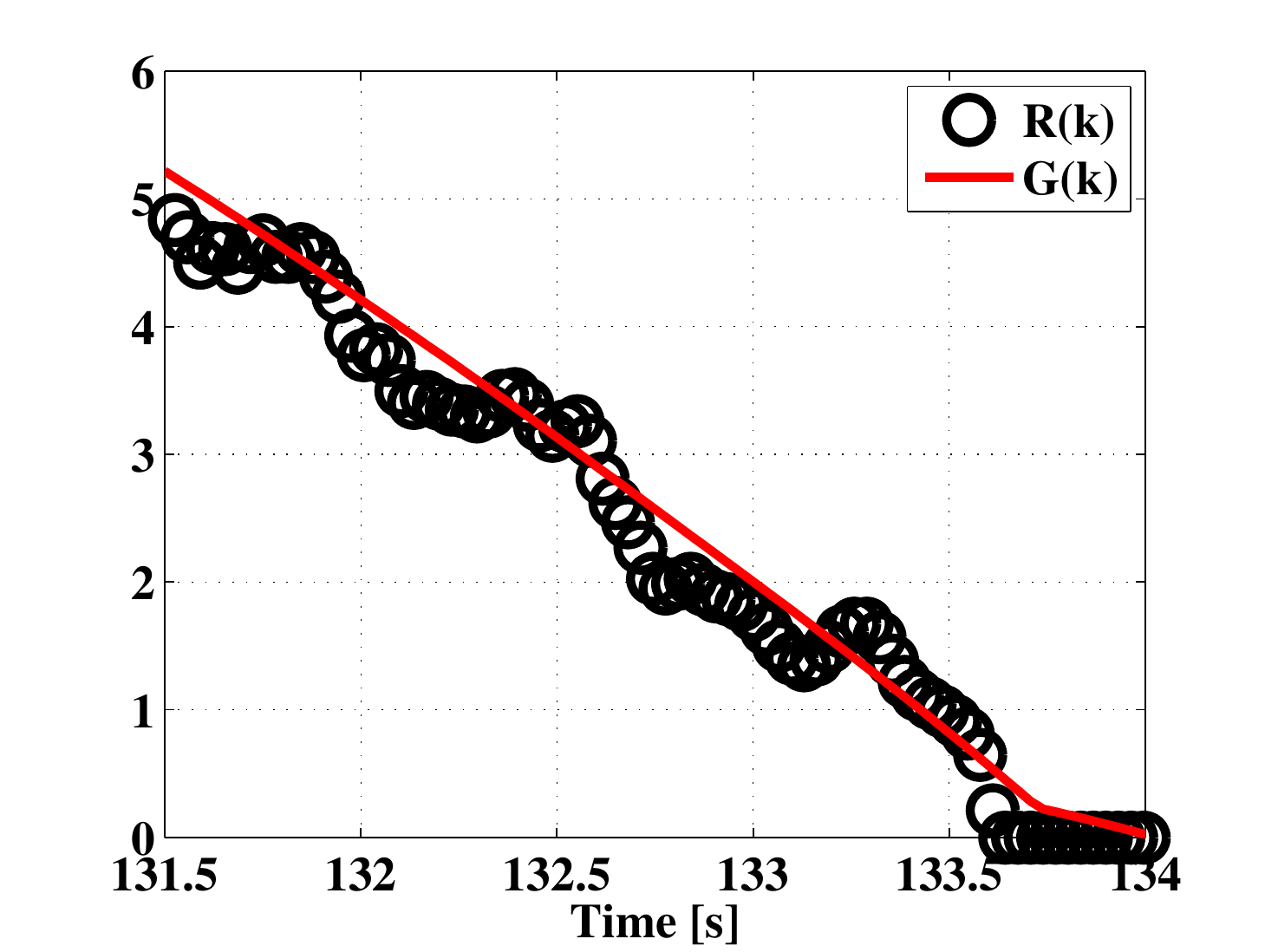}\label{fig:frequency_domain}}

\end{tabular}
\caption{Time and frequency domain comparison of modeled and measured RSS} 
\label{fig:models_directional}
\end{centering}
\end{figure*}

Figure~\ref{fig:freznel_zones} illustrates the location of the person in half a second intervals as the person approaches the LoS of a single link in Experiment $2$. The time domain RSS changes are illustrated in Fig.~\ref{fig:time_domain} and the frequency domain measurements in Fig.~\ref{fig:frequency_domain}. At time instant $t = 131.5 \text{ s}$, the person is in the 6$^{\text{th}}$ Fresnel zone and moving closer to the LoS. As shown in Fig.~\ref{fig:time_domain}, when the person approaches the LoS, they cause destructive and constructive fading sequentially as excess path length of the reflected signal changes. For example, at $t \approx 132.5 \text{ s}$ there is a constructive maxima and at this time instant, $\Delta = 2\lambda$ (third cross from the left in Fig.~\ref{fig:freznel_zones}). Thus, the single-bounce reflection model captures the RSS variations accurately with respect to the person's position.

As shown in Fig.~\ref{fig:frequency_domain}, also the modeled and measured frequency domain measurements correspond closely to each other. The true velocity of the person is modeled constant in the experiments resulting that the modeled frequency decays smoothly as the person approaches the LoS. Interestingly, the measured frequency varies around the modeled because it captures the small accelerations and deceleration of the movement. When a person walks, the velocity is not constant. Rather, it fluctuates around a certain mean with small variation around it as the person is walking. Moreover, it is to be noted that the rate of change decreases as the person moves closer to the LoS because it requires a larger spatial change of the person in the Cartesian coordinate system to alter $\Delta_r(k)$ the corresponding amount. The Fresnel zones in Fig.~\ref{fig:freznel_zones} illustrate this to some extent as they get sparser in the close vicinity of the LoS.

\subsection{Tracking Results}\label{S:tracking_results}

\begin{table*}[t]
    \caption{Results with and without frequency domain measurements when using $\check{v} \sim \mathcal{U}(0,2)$ and $\check{\theta} \sim \mathcal{U}(\theta - \frac{\pi}{4},\theta + \frac{\pi}{4})$ as initial estimates} 
        \centering 
	\renewcommand{\arraystretch}{1.2}
        \begin{tabular}{| c || >{\centering\arraybackslash}m{1.8cm} | >{\centering\arraybackslash}m{1.8cm} | 
			       >{\centering\arraybackslash}m{1.8cm} | >{\centering\arraybackslash}m{1.8cm} |
			       >{\centering\arraybackslash}m{1.8cm} | >{\centering\arraybackslash}m{1.8cm} |} 
	\hline\hline
	\multirow{2}{*}{Experiment} & \multicolumn{2}{c|}{$\bar{\epsilon}_x \pm \sigma_x$ [cm]} 
			            & \multicolumn{2}{c|}{$\bar{\epsilon}_y \pm \sigma_y$ [cm]}
			            & \multicolumn{2}{c|}{$\bar{\epsilon}_\%$} \\
				   
	\cline{2-7}
	& with $R(k)$ & w/o $R(k)$ & with $R(k)$ & w/o $R(k)$ & with $R(k)$ & w/o $R(k)$ \\
	\hline
	Ex. $1$ & $3.05 \pm 4.45$ & $9.01 \pm 19.32$ 
	        & $7.20 \pm 10.66$ & $20.52 \pm 31.34$ 
	        & $83.24$ & $51.31$ \\

	Ex. $2$ & $2.66 \pm 4.53$ & $9.76 \pm 25.98$ 
	        & $10.32 \pm 17.99$ & $24.21 \pm 39.15$ 
	        & $85.87$ & $54.36$ \\
	
	Ex. $3$ & $2.92 \pm 4.89$ & $10.50 \pm 26.62$ 
	        & $12.79 \pm 23.17$ & $27.48 \pm 43.55$ 
	        & $82.39$ & $51.80$ \\
			
	\hline
        \end{tabular}
        \label{table:tracking_accuracy1} 
\end{table*}

\begin{figure*}[!t]
\begin{centering}
\begin{tabular}{ccc}

\subfloat[$\varepsilon_x$]{\includegraphics[width=\columnwidth*30/100]{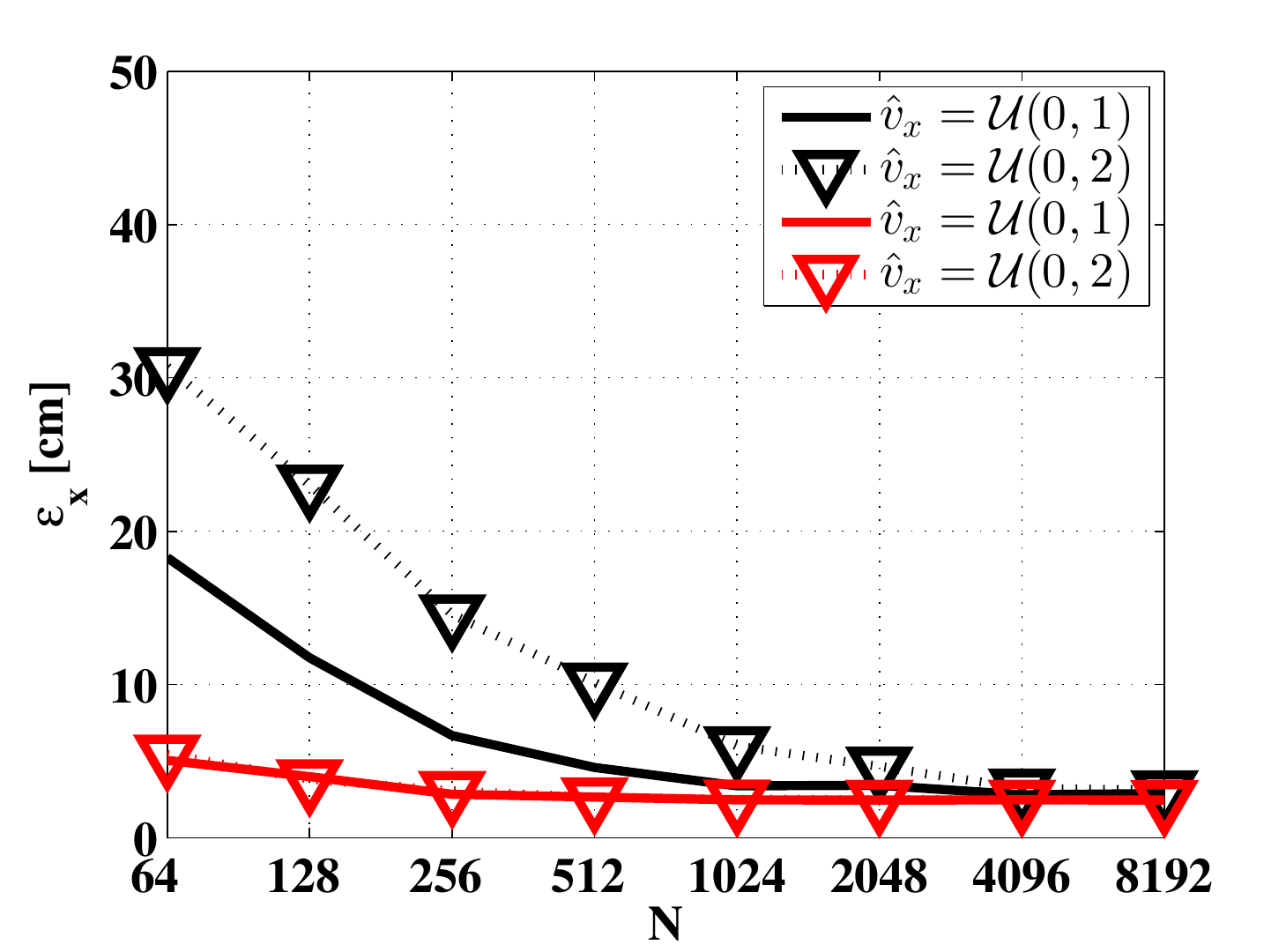}\label{fig:epsilon_x_N}}
\subfloat[$\varepsilon_y$]{\includegraphics[width=\columnwidth*30/100]{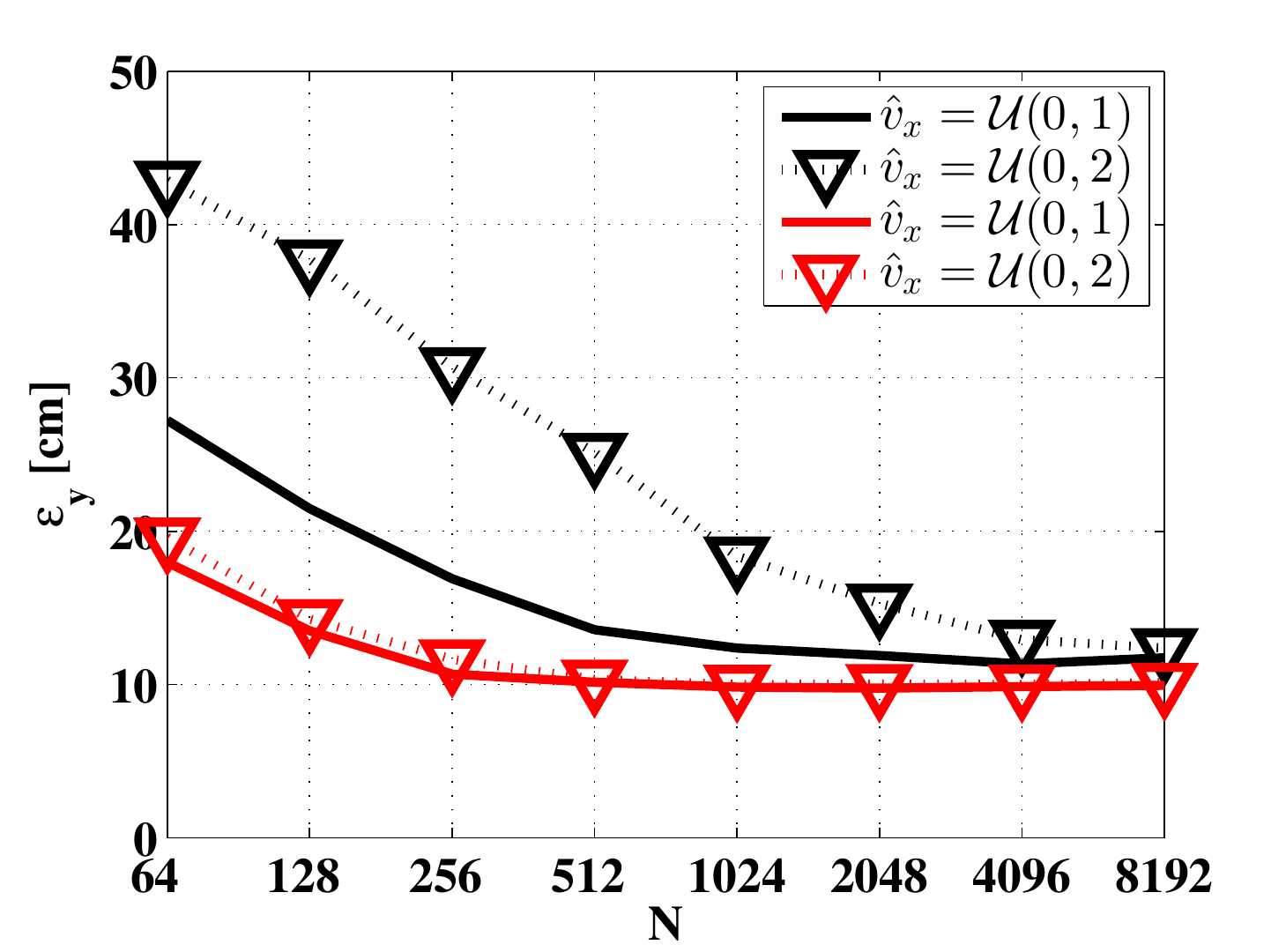}\label{fig:epsilon_y_N}}
\subfloat[$\varepsilon_\%$]{\includegraphics[width=\columnwidth*30/100]{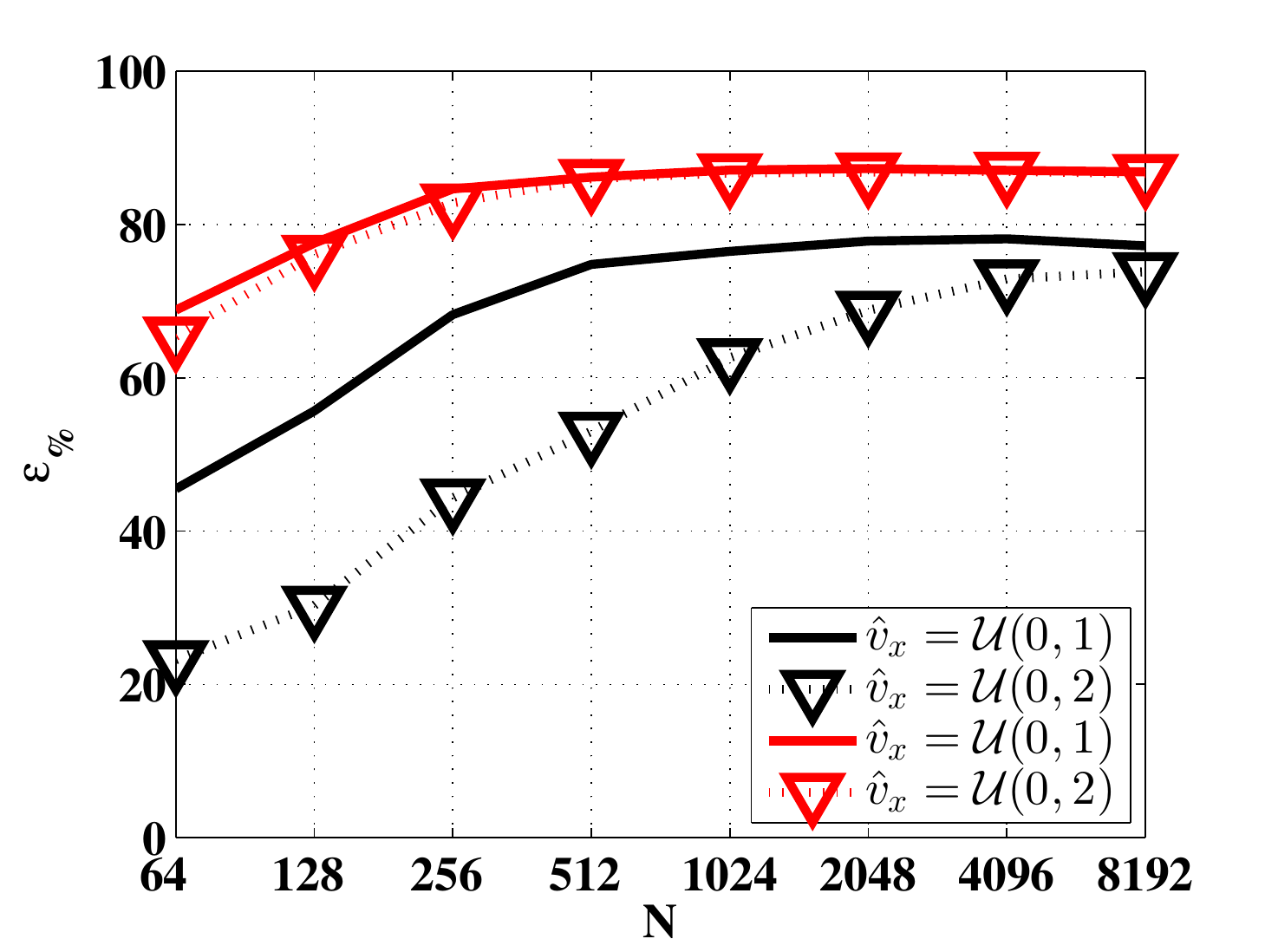}\label{fig:epsilon_N}}

\end{tabular}
\caption{Tracking performance with (red) and without (black) the frequency domain measurements as a function of particle number $N$}
\label{fig:particle_number}
\end{centering}
\end{figure*}

In Table \ref{table:tracking_accuracy1}, tracking accuracy of the system with and without the frequency domain measurements is given when the initial speed is initialized to $\check{v} \sim \mathcal{U}(0,2)$ and the heading to $\check{\theta} \sim \mathcal{U}(\theta - \frac{\pi}{4},\theta + \frac{\pi}{4})$. The accuracy of the system is approximately the same in the different experiments; only $\epsilon_y$ increases slightly with both systems due to the longer LoS distance which increases uncertainty of the y-coordinate. As shown in the table, the performance is enhanced by $58\%$ or more when the measurement model is augmented with the frequency domain measurements. Performance of the system that does not use the frequency domain measurements can be improved by increasing the number of used particles. Using $2048$ particles, $\epsilon_\% = \left[63.77 \% \quad   68.66 \% \quad 64.75 \% \right]$ for Experiment $1$, $2$ and $3$ in corresponding order. 

In Fig.~\ref{fig:particle_number}, the tracking performance in Experiment $2$ is illustrated as a function of particle number and with two different initial distributions for velocity. In the first initial distribution, speed is initialized to $\check{v} \sim \mathcal{U}(0,1)$ and the heading to $\check{\theta} \sim \mathcal{U}(\theta - \frac{\pi}{8},\theta + \frac{\pi}{8})$. In the second, uncertainty in the initial velocity estimate is increased an speed is initialized to $\check{v} \sim \mathcal{U}(0,2)$ whereas heading to $\check{\theta} \sim \mathcal{U}(\theta - \frac{\pi}{4},\theta + \frac{\pi}{4})$. The initial distribution of velocity has a considerably smaller effect on the system performance when the tracking algorithm is augmented with the frequency domain measurements. In fact, results using the two different initial distributions nearly coincide one another in this case. On the other hand, the initial distribution has a significant effect on the system performance when the system is solely relying on the time domain RSS measurements $-$ especially, when the number of particles is low. Without $R(k)$, the filter requires a large amount of particles to account for uncertainty in the initial estimate. This uncertainty can be considerably decreased when the propagation channel's change rate is estimated.

\begin{figure}[t]
\begin{centering}
\includegraphics[]{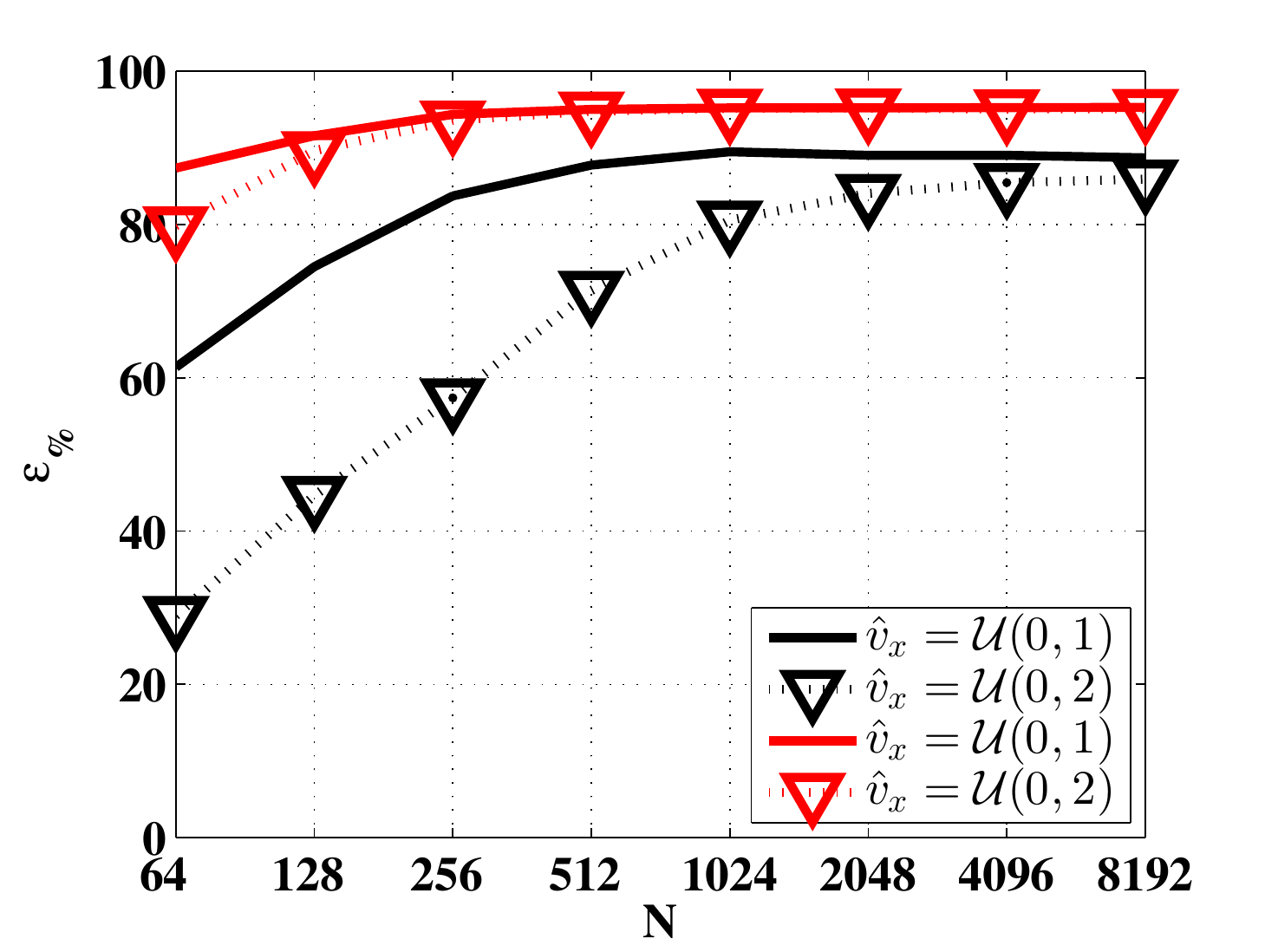}
\caption{Tracking performance with (red) and without (black) the frequency domain measurements when using three receivers}
\label{fig:three_receivers}
\end{centering}
\end{figure}

An additional receiver is deployed to the network in between the two receivers shown in Fig.~\ref{fig:experimental_setup} and Experiment $2$ is repeated. Using the three receivers, the tracking performance is illustrated in Fig.~\ref{fig:three_receivers} as a function of particle number. It is not surprising that the tracking accuracy is improved with both systems because the density of the network is increased and there are more measurements available for tracking. Comparing figures \ref{fig:epsilon_N} and \ref{fig:three_receivers}, one can observe that the results without the frequency domain measurements are enhanced more. However, using $R(k)$ the system achieves nearly perfect accuracy when $N \geq 256$ and there is no room for improvement. The results imply that systems that do not use the frequency domain measurements are more reliant on dense deployments, whereas systems using the frequency content of the measurements can also function in sparse deployments. This is especially desirable to decrease overall cost of RSS-based DFL systems.

In this paper, a standard laptop equipped with a $2.67$ GHz Intel Core i7-M620 processor and $8.0$ of GB of RAM memory is used. On average, computation time per iteration of the particle filter is $2.20 \text{ ms}$ without and $2.30 \text{ ms}$ with $R(k)$ when $N=512$. Thus, both systems can be easily implemented online since $T_s = 32 \text{ ms}$. Augmenting the system with the frequency domain measurements increases the computational demand by: i) calculating the frequency content of $\{r(k)\}_{k=1}^{N_f}$; ii) evaluating the two-dimensional measurement noise distribution in Eq.~\eqref{eq:residual_density}; iii) calculating $\frac{d}{d t}\Delta_r(k)$ in Eq.~\eqref{eq:rate_of_change}. All of these operations can be calculated efficiently and therefore, the computational demand only increases slightly when the frequency domain measurements are plugged into the tracking algorithm. 

\section{Discussion} \label{S:discussion}

\begin{figure*}[!t]
\begin{centering}
\begin{tabular}{cc}

\subfloat[Time domain]{\includegraphics[width=\columnwidth*4/10]{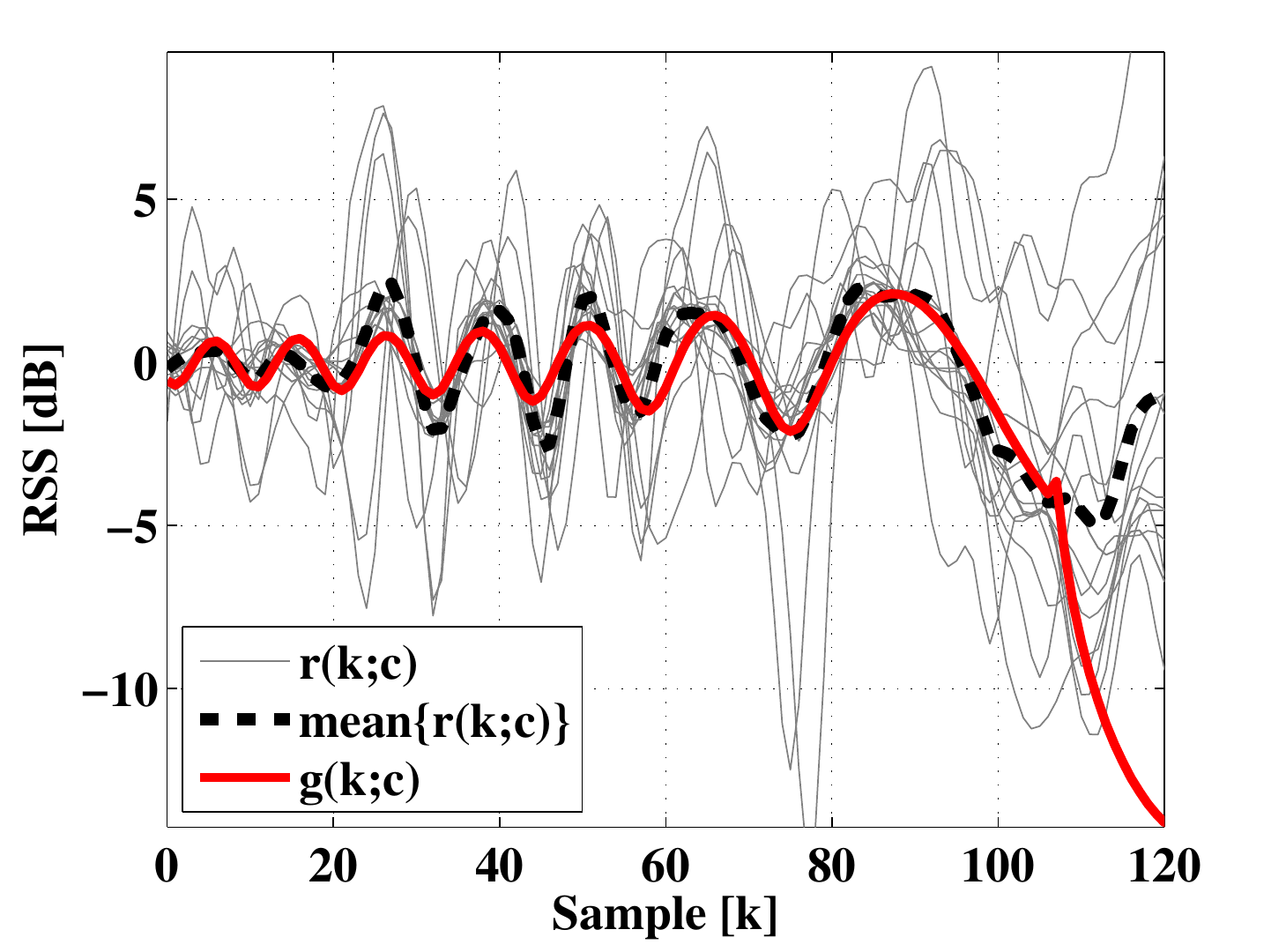}\label{fig:rss_omni_directional}}
\subfloat[Frequency domain]{\includegraphics[width=\columnwidth*4/10]{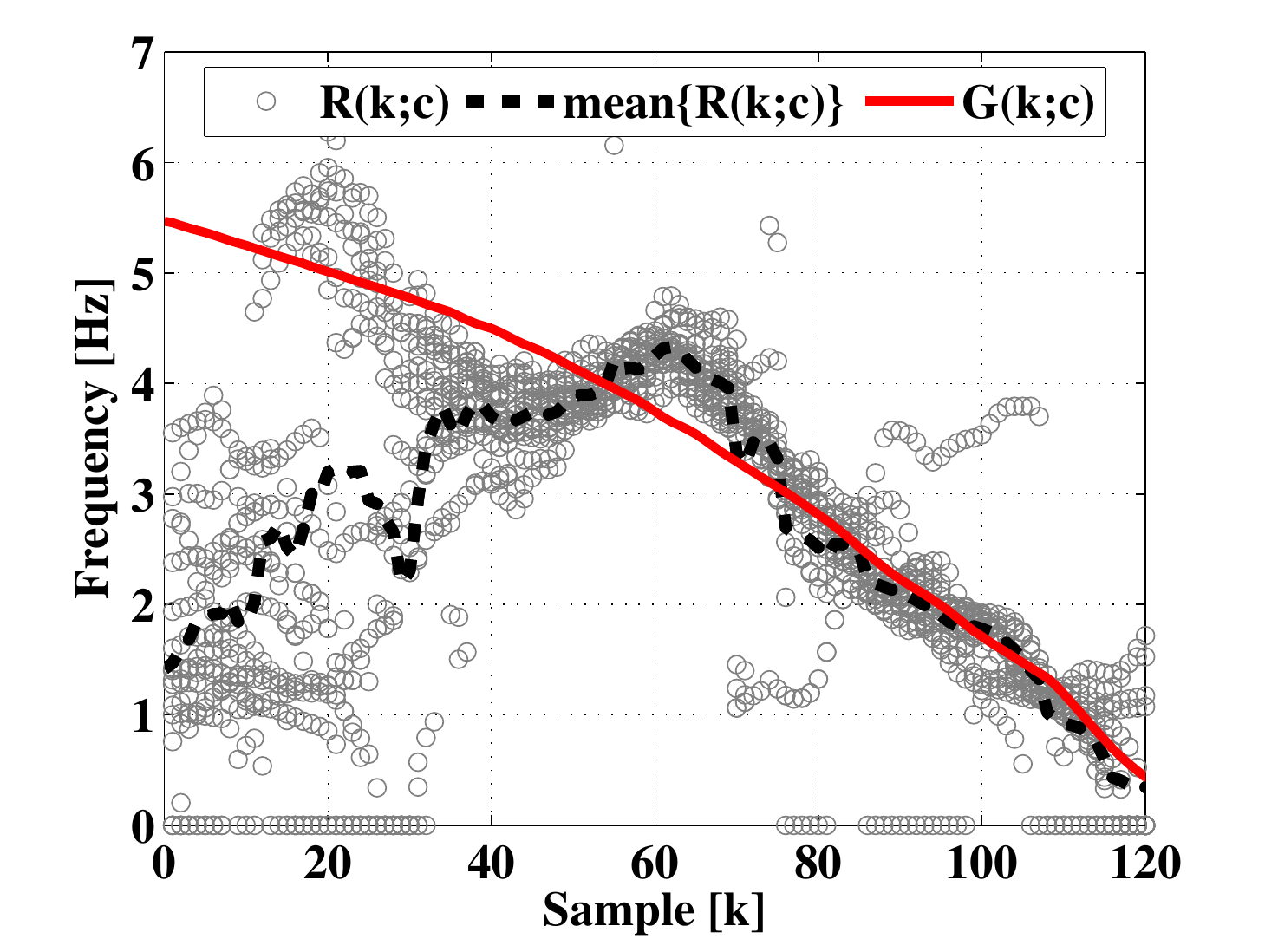}\label{fig:frequency_omni_directional}} 

\end{tabular}
\caption{Time and frequency domain comparison of modeled and measured RSS using omni-directional antennas}
\label{fig:omni_directional}
\end{centering}
\end{figure*}

In this paper, using the frequency domain measurements for tracking is demonstrated using directional antennas. For comparison, RSS measurements on $16$ different frequency channels using omni-directional antennas are shown in Fig.~\ref{fig:omni_directional} for the deployment scenario of Experiment $2$. In more cluttered environments and with omni-directional antennas, the time domain measurements tend to deviate from the modeled because of multipath propagation. Despite the fact that some of the frequency channels do not correspond the modeled RSS as shown in Fig.~\ref{fig:rss_omni_directional}, still, the frequency domain measurements are close to the modeled when the person is relatively close the LoS, i.e., $k \geq 40$. Thus, it can be concluded that the frequency domain measurements are not as easily altered by noise, measurement equipment induced effects such as carrier and symbol synchronization, or electrical properties of the person. As a consequence, we anticipate that exploiting the frequency domain measurements is especially beneficial when using omni-directional antennas and in environments where RSS-based DFL is difficult, i.e., in through-wall scenarios and in cluttered indoor environments. We will explore this possibility in future research. 

Using frequency domain measurements for tracking sets stringent requirements on sampling frequency of the wireless channel. In this paper, we have demonstrated tracking when a person is moving at a speed of $0.5 \text{ m/s}$ and shown that with such a speed, the propagation channel's change rate is approximately $6 \text{ Hz}$ or below. Thus, the channel should be sampled at $12 \text{ Hz}$ or above to satisfy the Nyquist rate. This sampling frequency is hard to accomplish in mesh networks where all transceivers are communicating with one another. Moreover, exploiting channel diversity makes the task even more challenging since it increases the sampling rate with respect to the number of used channels. There are three options one can investigate to overcome this limitation. First and most importantly, the experiments are conducted with hardware that is designed for low-rate and low-power communications. Using customized or alternative technologies such as Bluetooth could be beneficial for RSS-based DFL since the propagation medium could be sampled at a higher rate and frequency hopping would be part of the protocol implementation. Second, the communication schedule could be designed adaptive based on the location of the person. In this case, only links that traverse near by the person's location would participate in the localization effort. Third, compressed sensing could be used to reconstruct the signal of interest from undersampled RSS measurements.

\section{Conclusions} \label{S:conclusions}
In this paper, frequency content of the RSS measurements is shown to contain information about a person's velocity. A closed form expression that relates the frequency content of the RSS measurements to the person's position and velocity is derived. The system model is augmented with the additional measurement and a particle filter is implemented to track movements of a person. Simulations are used to evaluate the system performance and experiments are conducted to validate the development efforts. The results suggest that the tracking accuracy is improved while the system's robustness to parameter changes is increased when the frequency domain measurements are plugged into the tracking algorithm. As discussed in the paper, the frequency domain measurements are a robust measurement metric for localization purposes and we anticipate that using them is especially beneficial with systems relying on omni-directional antennas and in environments where RSS-based DFL is difficult, i.e., in through-wall scenarios and in cluttered indoor environments.

\end{document}